# Change of Cropping Patterns of Southeastern Anatolia, Turkey in 2019 and 2020

Mustafa Teke

mustafa.teke@tubitak.gov.tr

TÜBİTAK UZAY Space Technologies Research Institute

Ankara, Turkey

Field crops are a major freshwater consumer. Crop mapping from satellite imagery may help estimation of water usage. In this short report, change of cropping practices between 2019 and 2020 will be described by using remote sensing and machine learning methods.

Analysis were performed in the Southeastern Anatolia Region of Turkey.

## Data

Analysis were performed by using time series Sentinel-2 imagery. Data was download from Sentinel-Hub. Later processed by FMask V4 to produce cloud and shadow masks.

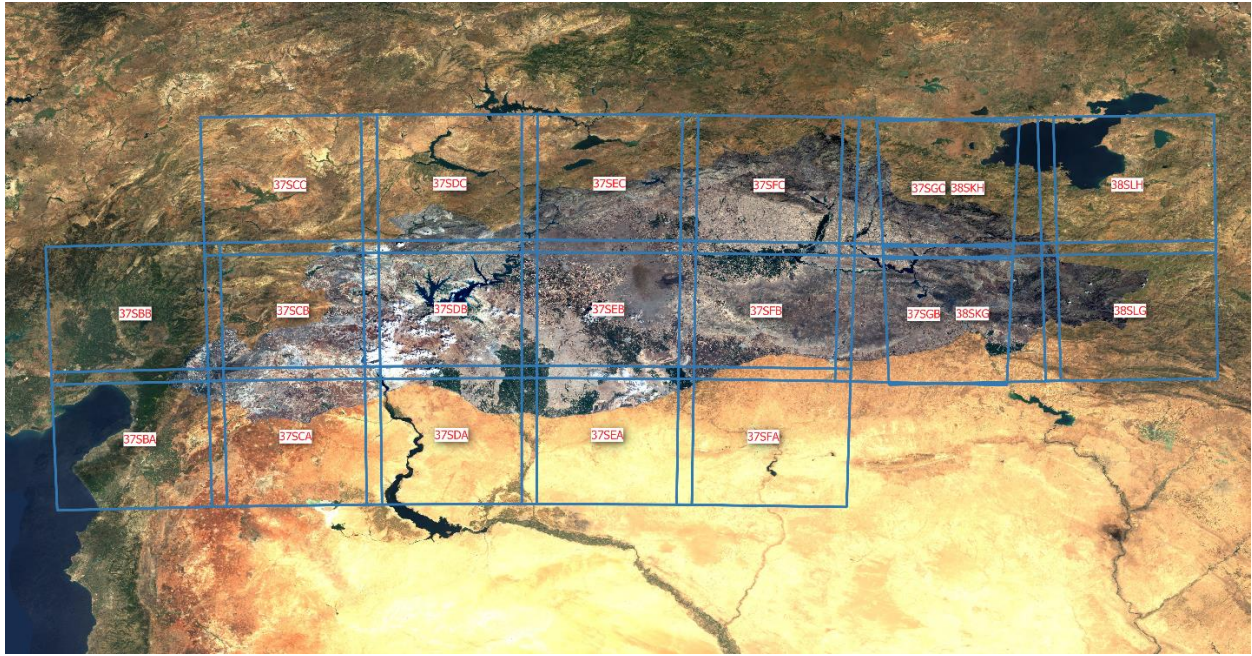

*Figure 1 Sentinel-2Tiles used in the analysis.*

Potential crop areas were selected by CORINE 2018 *[1]* data to reduce processing times. These areas are displayed in Figure 2.

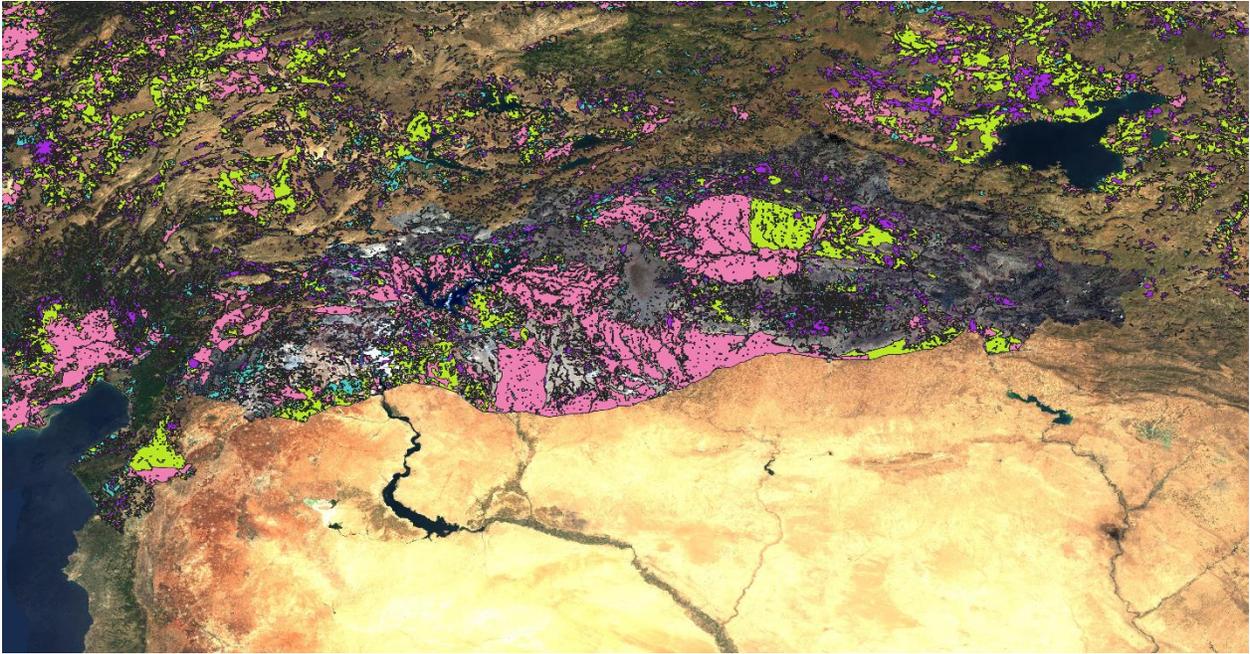

*Figure 2 Agricultural Areas of CORINE 2018 dataset. Lands are classified according to their irrigation status.*

## Method

Vector Dynamic Time Warping (VDTW) method was used in classification with spectral signatures of major crops in the region [2]. Grains include wheat, barley, lentil, chickpea and Fig. Phenological properties of crops of interest are shown below:

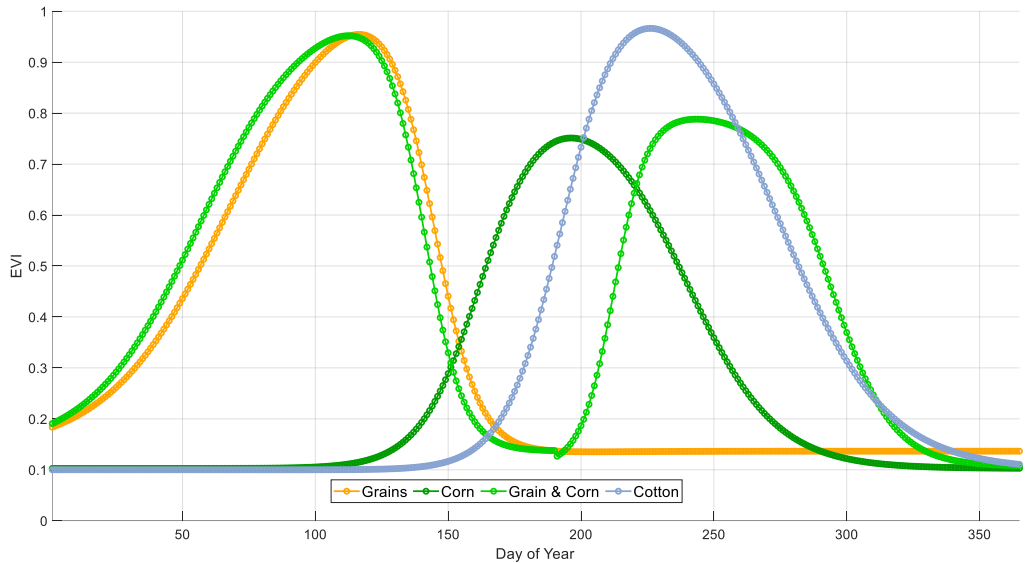

*Figure 3 Phenologies of major crops in Southeastern Anatolia Region.*

Enhanced Vegetation Index (EVI) was used as the phenology indicator since it provides higher discrimination between different crops compared to Normalized Difference Vegetation Index (NDVI).

## Results

Our crop mapping maps for 2019 and 2020 are presented in Figure 4 and Figure 5 respectively. Corn and double cropping corn are merged.

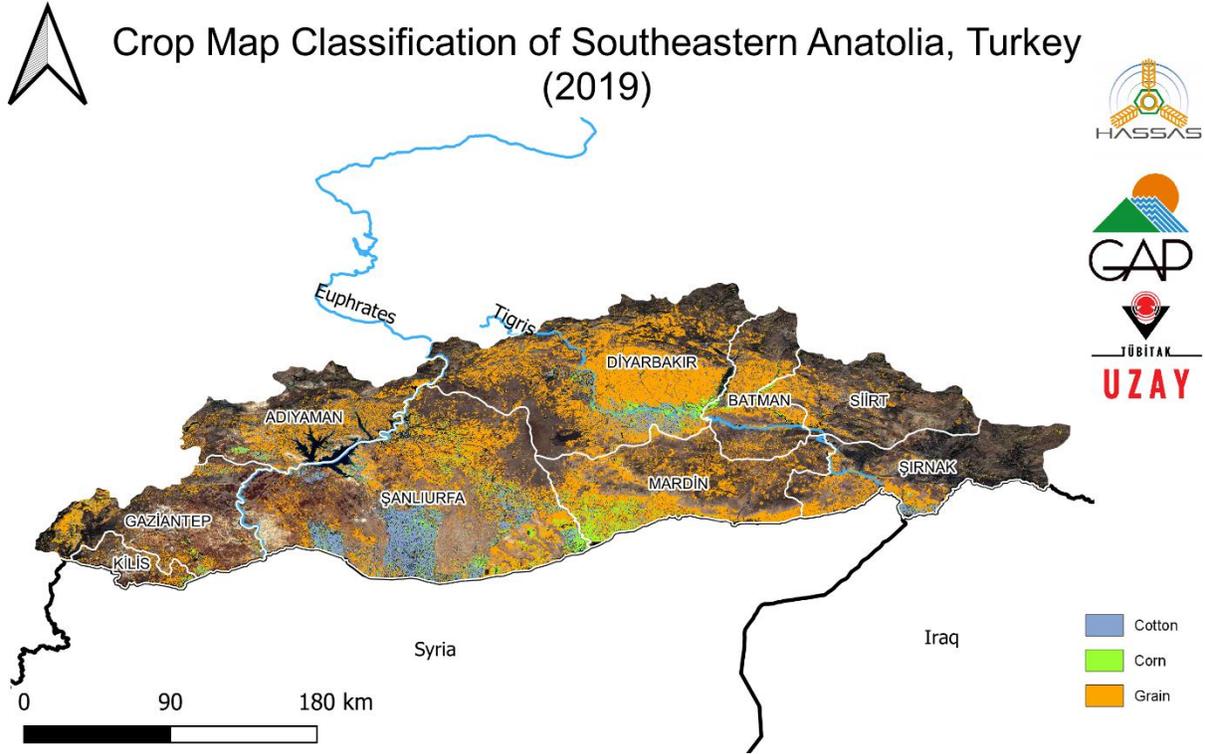

*Figure 4 Southeastern Anatolia Region 2019 Major Crops.*

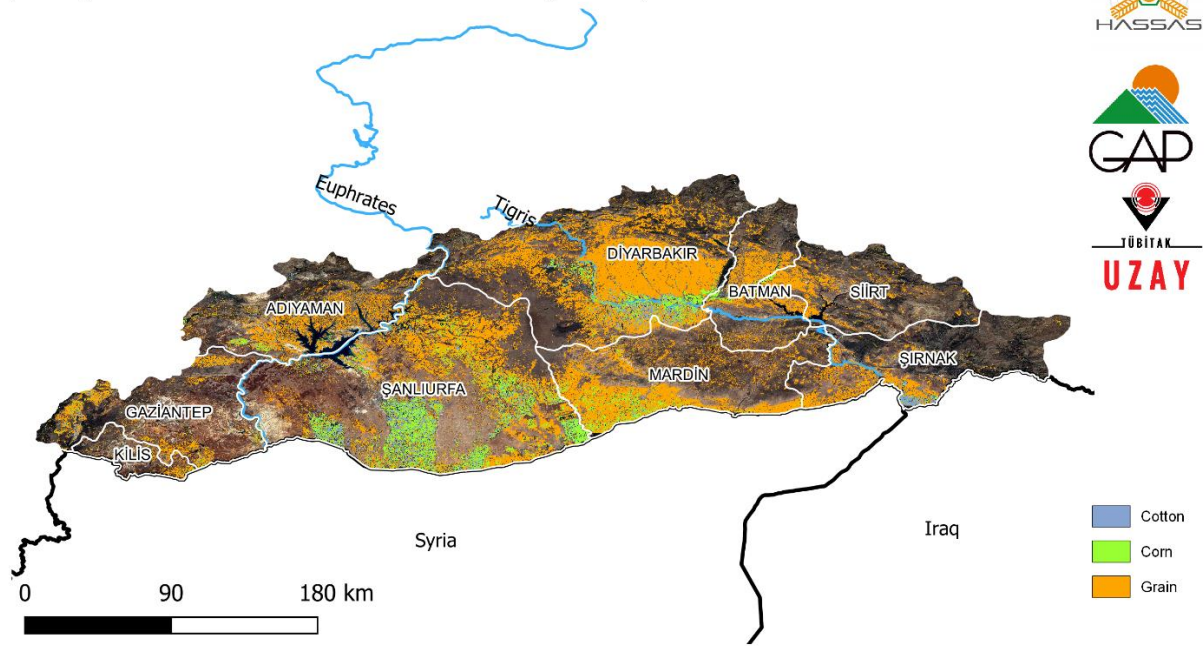

*Figure 5 Southeastern Anatolia Region 2020 Major Crops.*

Grains are major crop type in the region. Cotton was common in 2019 however this trend changed in 2020 in favor of Corn.

Change of cropping practices between 2013 and 2020 is available in [3].

## Conclusions

Time series satellite imagery is useful in multi-year analysis in crop growing practices. Climate, economic and government subventions are main factors effecting crops grown by the farmers.

The region is fed by to major rivers Tigris and Euphrates and region is irrigated by Southeastern Anatolian Project.

These crop maps may be used to understand water use planning in the region since agricultural activities requires most of the fresh water.

## Acknowledgements

This study is supported under Widespread application of sustainable precision agriculture practices in Southeastern Anatolia Project Region (GAP) by Ministry of Industry and Technology, Southeastern Anatolia Project Regional Development Administration.

## References

[1]     Copernicus Land Monitoring Service, "CLC 2018," 2020. https://land.copernicus.eu/pan-european/corine-land-cover/clc2018.

[2]  M. Teke and Y. Yardımcı, "Multi-Year Vector Dynamic Time Warping Based Crop Mapping," *arXiv*. 2019.

[3]  M. Teke, "Harran Ovası Ürün Deseni Değişimi (Turkish)," 2020. https://medium.com/@mustafa.teke/harran-ovası-ürün-deseni-değişimi-a75319140518.